# High-frequency performance of scaled carbon nanotube array field-effect transistors

Mathias Steiner[1,a], Michael Engel[2,3], Yu-Ming Lin[1], Yanqing Wu[1], Keith Jenkins[1], Damon B. Farmer[1], Jefford J. Humes[4], Nathan L. Yoder[4], Jung-Woo T. Seo[5], Alexander A. Green[5], Mark C. Hersam[5], Ralph Krupke[2,3,6], and Phaedon Avouris[1,b]

[1] *IBM Thomas J. Watson Research Center, Yorktown Heights, New York 10598, USA*

[2] *Institute of Nanotechnology, Karlsruhe Institute of Technology, 76021 Karlsruhe, Germany*

[3] *DFG Center for Functional Nanostructures (CFN), 76028 Karlsruhe, Germany*

[4] *NanoIntegris Inc., Skokie, IL 60077, USA*

[5] *Department of Materials Science and Engineering and Department of Chemistry, Northwestern University, Evanston, IL 60208, USA*

[6] *Institut für Materialwissenschaft, Technische Universität Darmstadt, 64287 Darmstadt, Germany*

---

[a] Electronic mail: msteine@us.ibm.com
[b] Electronic mail: avouris@us.ibm.com

**Abstract:**

We report the radio-frequency performance of carbon nanotube array transistors that have been realized through the aligned assembly of highly separated, semiconducting carbon nanotubes on a fully scalable device platform. At a gate length of 100 nm, we observe output current saturation and obtain as-measured, *extrinsic* current gain and power gain cut-off frequencies, respectively, of 7 GHz and 15 GHz. While the extrinsic current gain is comparable to the state-of-the-art the extrinsic power gain is improved. The de-embedded, *intrinsic* current gain and power gain cut-off frequencies of 153 GHz and 30 GHz are the highest values experimentally achieved to date. We analyze the consistency of DC and AC performance parameters and discuss the requirements for future applications of carbon nanotube array transistors in high-frequency electronics.

Field-effect transistors made of semiconducting carbon nanotubes (CNTs) have excellent electrical DC characteristics and have been explored as possible successors to their silicon counterparts[1]. Knowledge of their AC characteristics is, however, limited. For transistors made of a single CNT, because of the large impedance mismatch with the measurement system, such studies are extremely difficult[2,3,4,5,6,7] and, ultimately, their output current is not sufficient for technological applications. In order to deploy CNTs in high-frequency electronics it is necessary to use an array[8,9,10,11,12]. Nevertheless, there are problems associated with their separation, their aligned assembly at high densities, scaling of device dimensions, and minimization of device parasitics[13].

In this Letter, we introduce a scalable, planar device platform that enables the electric-field driven, *in situ* assembly of aligned CNT arrays from solution. We characterize field-effect transistors made of those arrays by means of electrical transport measurements in the DC and AC domain. Furthermore, we investigate the consistency of the experimental results and discuss the requirements for future applications of CNT array transistors in high-frequency electronics.

We produced CNT array field-effect transistors based on centrifugation-enriched solutions containing 99.6% semiconducting carbon nanotubes having an average diameter of 1.5nm[14]. A planar device platform with embedded electrodes allows for the controlled scaling of channel length, gate length, gate dielectric thickness, and contact length (see Fig. 1). Electron beam lithography with poly(methyl methacrylate) (PMMA) as the resist was used for all the patterning. In a first step, the device layout and alignment markers were patterned into the PMMA on top of highly resistive Si coated with a $SiO_2$ layer having a thickness of 1μm, see Fig. 1(a). The pattern was transferred into the underlying $SiO_2$ with a combination of reactive ion etching and wet chemical etching using buffered oxide etch, see Fig. 1(b). The resulting etch depth was 50 nm. The metal stack consisting of 5nm Ti, 45nm Au, and 1nm Ti was then sequentially evaporated on top of the structure. After lift-off in acetone, we obtained a planar device platform with embedded metal electrodes as shown in Fig. 1(c), thus avoiding performance degradation due to bending of the CNTs. In the next step, a gate dielectric layer made of 10nm $HfO_2$ ($\kappa$=13) grown by atomic layer deposition at 125°C was locally deposited through a PMMA mask window followed by lift off processing, see Fig. 1(d). We then assembled aligned CNT arrays having a density of $D = 50\mu m^{-1}$ from an aqueous CNT solution by using low-frequency dielectrophoresis[15]: a droplet (~30μl) of diluted CNT suspension (0.5μg/ml) was placed on the device structure while an alternating voltage ($V_{pp} = 4V, f = 300kHz$) was applied between the deposition electrode pairs for 10 minutes as schematically indicated in Fig. 1(e). The CNT solution was then replaced with de-ionized water and the surface was dried with a stream of $N_2$, see Fig. 1(f). We note that the CNTs do not make physical contact with the deposition electrodes underneath the dielectric. This way, material issues associated with heat dissipation during dielectrophoresis, such as electro-migration, are avoided. The device fabrication was completed by patterning source and drain contacts and evaporation of 0.5nm Ti, 40nm Pd, and 20nm Au, see Fig. 1(g). The (first-layer) deposition electrodes below the gate dielectric layer and the actual (second-layer) source and drain contacts of the device were shorted in order to reduce parasitic capacitances. Scanning electron microscopy images of the CNT array transistor are shown in Fig. 1(h).

Fig. 2 shows the DC electrical characterization of the CNT array transistor. The device exhibits a current density of $I_d = 50\mu A/\mu m$ and a transconductance of $g_m = \Delta I_d/\Delta V_g = 15 \mu S/\mu m$ at a gate length of $L_g = 100nm$. Most importantly, current saturation is observed in the output characteristics of the device; see Fig. 2(b). Accordingly, the output conductance $g_d = \Delta I_d/\Delta V_d$ drops to zero at a drain bias of $V_d = -2V$.

We determined the high-frequency performance of the CNT array transistors by measuring the scattering (S) parameters (see supplementary material). From this, we obtained the short circuit current gain cut-off frequency ($f_T$) and the maximum frequency of oscillation ($f_{MAX}$) as shown in Fig. 3(a). These frequencies characterize the highest frequencies at which electrical signals are propagated in the transistor and electrical power gain is achieved, respectively. The as-measured, extrinsic current gain $|h_{21}|$ of the CNT array transistor shows the expected -20dB/decade roll-off with an extrinsic cut-off frequency of $f_T^{ext} = 7GHz$ which is comparable to the state-of-the-art[10,12]. Both the extrinsic unilateral power gain (U) and the extrinsic maximum available power gain (MAG) become 0dB at $f_{MAX}^{ext} = 15GHz$ which constitutes an improvement with respect to the state-of-the-art[10,12].

The performance of scaled high-frequency transistors is typically deteriorated by parasitic resistances and parasitic capacitances[16]. While the extrinsic AC performance of a transistor is meaningful in the context of circuit design and integration, the intrinsic AC performance reveals the electrical transport properties of the channel material itself. The determination of the intrinsic transistor performance is routinely accomplished by de-embedding techniques[17,18] that have been widely adopted for characterizing electronic devices made of bulk semiconductors, as well as low-dimensional materials. The intrinsic AC cut-off frequency $f_T^{int}$ is proportional to the DC transconductance $g_m$ of the transistor and can be expressed by[13]:

$$f_T^{int} = \frac{g_m}{2\pi C_g}.\tag{1}$$

This expression allows for testing the consistency between the AC and DC measurements and constitutes an appropriate figure of merit to compare the performance potential of different semiconducting channel materials for high-frequency applications[13].

We use a de-embedding procedure based on "open" and "short" reference structures to determine the intrinsic high-frequency performance of the CNT array transistor (see supplementary material). Based on the extrinsic results shown in Fig. 3(a), we obtain an intrinsic current gain cut-off frequency $f_T^{int} = 153\text{GHz}$ and an intrinsic power gain cut-off frequency $f_{MAX}^{int} = 30\text{GHz}$ for $L_g = 100\text{nm}$, see Fig. 3(b). The results demonstrate that the advancements made with respect to CNT separation, alignment and device scaling result in the highest intrinsic high-frequency CNT transistor performance reported so far.

We now investigate the consistency of the electrical transport measurements (DC and AC) and calculate the theoretical gate capacitance $C_g^{theo}$ of an evenly spaced array of CNTs with density $D$, which is given by[19]:

$$C_g^{theo} = D\left\{\frac{1}{2\pi\varepsilon_0\varepsilon_r}\ln\left[\frac{\sinh(2\pi tD)}{\pi rD}\right] + \frac{1}{C_q}\right\}^{-1},\tag{2}$$

where $C_Q = 4\cdot 10^{-10}\,\text{F}\cdot\text{m}^{-1}$ is the quantum capacitance of a CNT and $r = 0.75\text{nm}$ is the average CNT radius. As in our experiments, the gate dielectric is HfO$_2$ having a thickness $t = 10\text{nm}$ and we assume a relative permittivity $\varepsilon_r = (13+1)/2$ to account for the fact that the CNTs are located at the HfO$_2$/air interface. By using a CNT density of $D = 50\mu\text{m}^{-1}$, we obtain $C_g^{theo} = 0.0013\text{pF}$. In Fig. 4(a), we plot the experimental $f_T^{int}$-values as function of the associated $g_m$-values for four different devices with $L_g = 100\text{nm}$. By fitting Eq. (1) to the data, we obtain an experimental gate capacitance of $C_g^{exp} = (0.0013 \pm 0.0001)\text{pF}$ which is in agreement with the calculated $C_g^{theo}$-value. Furthermore, by using the experimental DC-value $g_m = 15\mu\text{S}/\mu\text{m}$ and $C_g^{exp} = C_g^{theo} = 0.0013\text{pF}$ in Eq. (1), we obtain $f_T^{int} = 153\text{GHz}$, which is in agreement with the result of the S-parameter measurement, see Fig. 3(b).

We now analyze the intrinsic power gain cut-off frequency $f_{MAX}^{int}$ which we relate to $f_T^{int}$ in the following way[13]:

$$f_{MAX}^{int} = \frac{f_T^{int}}{2\sqrt{g_d(R_{p,s} + R_g) + 2\pi f_T^{int} C_{p,gd} R_g}}. \qquad (3)$$

Here, $R_{p,s}$ is the parasitic source resistance ($R_{p,s} \approx 0.5 R_{device} = 250\Omega$ in the present case), $R_g = 45\Omega$ is the gate resistance, and $C_{p,gd} \approx \frac{2}{3} \cdot C_{gate} = 0.03\text{pF}$ is the parasitic gate-to-drain capacitance that is estimated based on the geometric gate capacitance of the device, $C_{gate}$. By using the experimental values $g_d = 25\mu\text{S}/\mu\text{m}$ and $f_T^{int} = 153\text{GHz}$ in Eq. (3), we obtain $f_{MAX}^{int} = 55\text{GHz}$. In the case where we use $C_{p,gd} = 0.13\text{pF}$, we are able to reproduce the experimental value $f_{MAX}^{int} = 30\text{GHz}$ obtained by the S-parameter measurement, see Fig. 3(b). This relatively high $C_{p,gd}$-value of 0.13pF can be understood by considering the presence of the deposition electrodes in the device (see Fig. 1) which are not accounted for in the conservative estimate $C_{p,gd} \approx \frac{2}{3} \cdot C_{gate}$. This analysis validates the consistency of the electrical transport measurements and shows the good correlation between the experimental AC parameters $f_T^{int}$ and $f_{MAX}^{int}$ and the measured DC parameters $g_m$ and $g_d$.

Finally, we discuss important technical requirements for future applications of CNT array transistors in high-frequency electronics. A key issue that needs to be addressed is the high contact resistance of CNTs in the solution-processed array. Based on our DC measurements, we estimate that the contact resistance of a single CNT can be as high as 500kΩ which is 50 times larger than the values achieved with single, as grown CNTs[20]. This is likely due to incomplete removal of residual surfactant surrounding the CNTs. Lowering the resistance of the CNT-metal contacts would benefit both intrinsic and extrinsic device performance.

Significant performance improvements could be realized through device design and scaling over the course of this study. In Fig. 4(b), we plot the maximum intrinsic $f_T$-values that we obtained with different device concepts while maintaining the quality of CNT solutions at a high level of electronic separation (<1% metallic carbon nanotube admixture). Shortcomings of our previous top-gated and back-gated device concepts involved limited scalability, excessive gate leakage, and performance-degrading chemical treatment and bending of CNTs. The device concept reported here has overcome those limitations and holds great potential for further scaling and performance improvements. Implementing the manufacturing principle on a flexible plastic substrate as discussed in reference 21 could enable high-frequency applications that are not possible with state-of-the-art thin film transistor technology.

Further enhancements in performance are expected in transistors made of CNT solutions having a higher selectivity with respect to their diameter and, ultimately, with solutions containing only a single CNT chiral species[22,23,24], which may allow the application of such devices in logic circuits. Nevertheless, operation in analog radiofrequency circuitry[8] is conceivable with the CNT solutions used in this study. In this respect, we note that our DC measurements indicate a self gain of the order $g_m/g_d \approx 10$ and further optimization is certainly possible. We point out, however, that the large-scale assembly of dense and aligned arrays of solution-processed CNTs remains a major challenge.

As compared to high-frequency transistors made of graphene[25], the presence of a band-gap in CNTs and the current saturation at shorter gate lengths suggest the possibility of higher power gain and lower standby power consumption.

**Acknowledgment**

We acknowledge discussions with A. D. Franklin, W. Haensch, S. O. Koswatta, and A. Valdes-Garcia (all IBM T. J. Watson Research Center) and thank J. J. Bucchignano, B. A. Ek, and G. P. Wright (IBM T. J. Watson Research Center) for expert technical assistance. M.C.H. acknowledges support by the National Science Foundation (DMR-1006391 and DMR-1121262) and the Nanoelectronics Research Initiative. N.L.Y. acknowledges support by the Office of Naval Research (ONR), grant # N00014-11-C-0135.

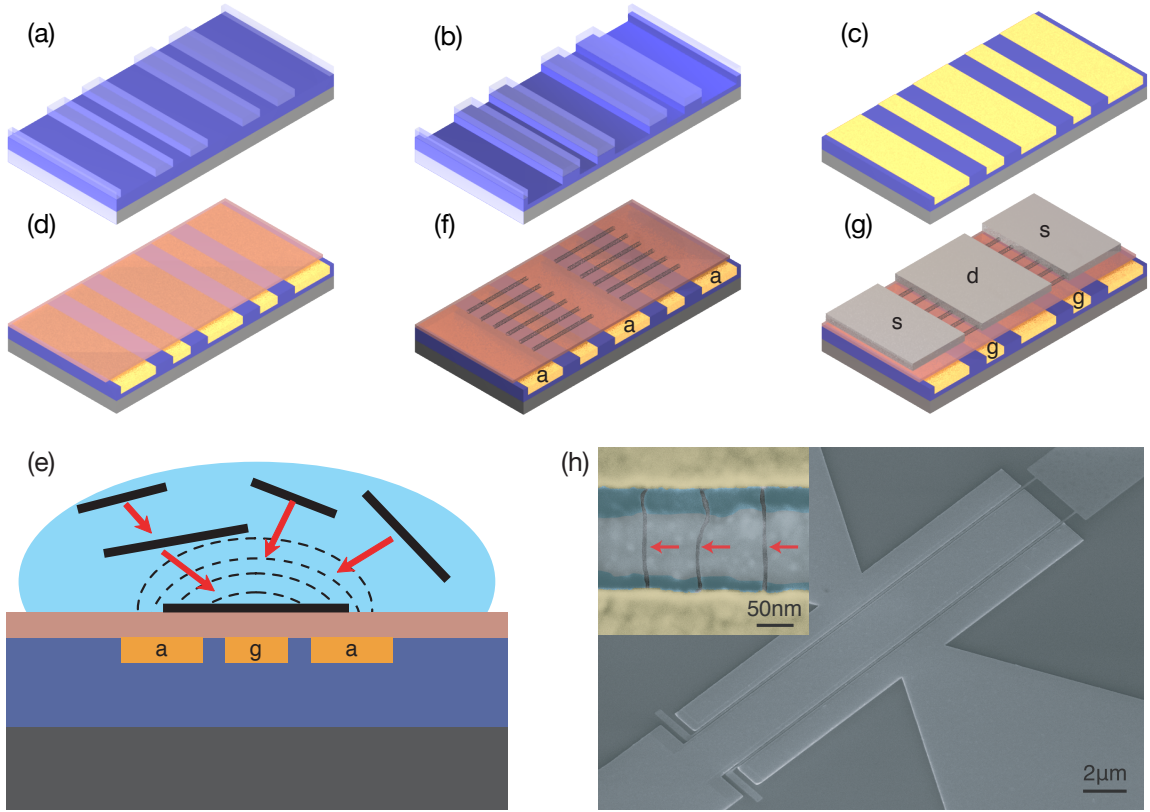

FIG. 1. Solution-assisted manufacturing of scaled carbon nanotube array transistors. (a-g) Schematic illustrations of the manufacturing steps for obtaining a planar device with embedded gate electrodes [labeled g in (e),(g)] and deposition electrodes [labeled a in (e),(f)]. In (g), source (s) and drain (d) electrodes of the transistor are also indicated. (h) Scanning electron microscope image of the dual channel device having a channel width of 2x20μm. (Inset) Magnified and colorized scanning electron microscopy image of the transistor channel showing 3 parallel carbon nanotubes (highlighted by red arrows) bridging the gated region (gray color) between source and drain electrodes (golden color).

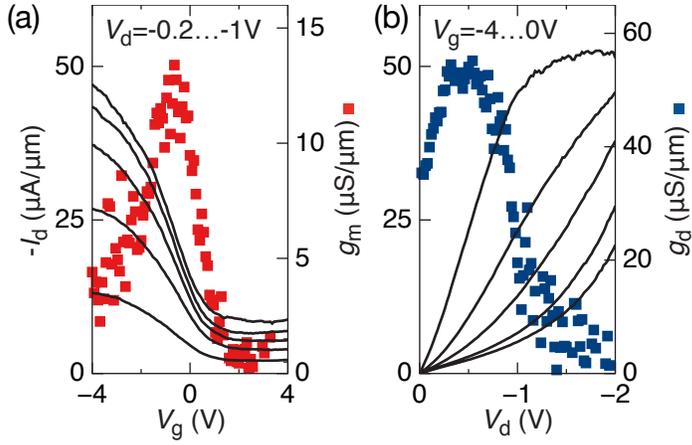

FIG. 2. Electrical DC characteristics of the carbon nanotube array transistor. (a) Measured electrical $I_d$-$V_g$ transfer characteristics (black lines) and transconductance $g_m$ (red squares). (b) Measured electrical $I_d$-$V_d$ output characteristics (black lines) and output conductance $g_d$ (blue squares). The magnitude of the electrical parameters $V_d$, $V_g$ in the respective voltage sweeps are indicated on top.

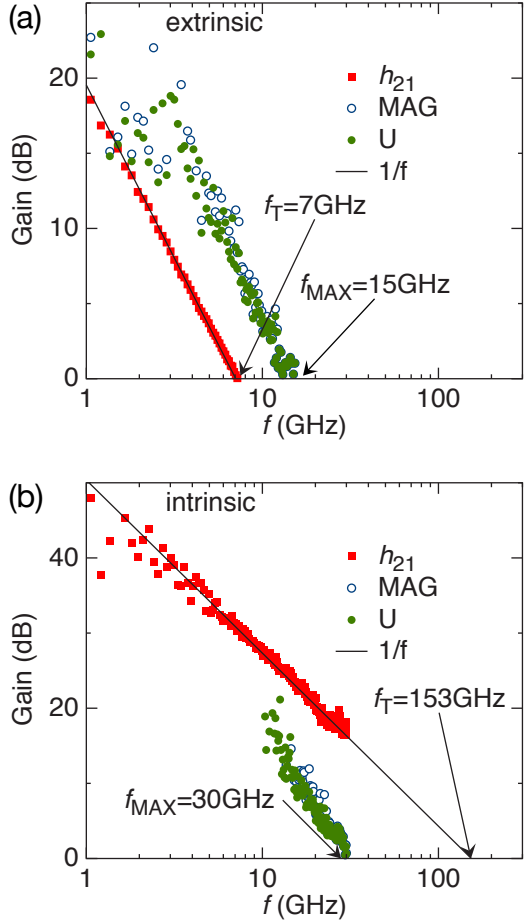

FIG. 3. Electrical AC characteristics of the carbon nanotube array transistor. (a) The as-measured, extrinsic short-circuit current gain ($h_{21}$, red squares) shows -20dB/decade roll-off (black line: $1/f$-fit) and becomes 0dB at the current gain cut-off frequency $f_T$. The extrinsic, maximum available power gain (MAG, blue rings) and the extrinsic, unilateral power gain (U, green circles) becomes 0dB at the maximum frequency of oscillation, $f_{MAX}$. (b) Intrinsic short-circuit current gain ($h_{21}$, red squares; black line: $1/f$-fit), intrinsic maximum available power gain (MAG, blue rings), and intrinsic unilateral power gain (U, green circles). The measurement was performed at $V_d$=-2V, $V_g$=-2V.

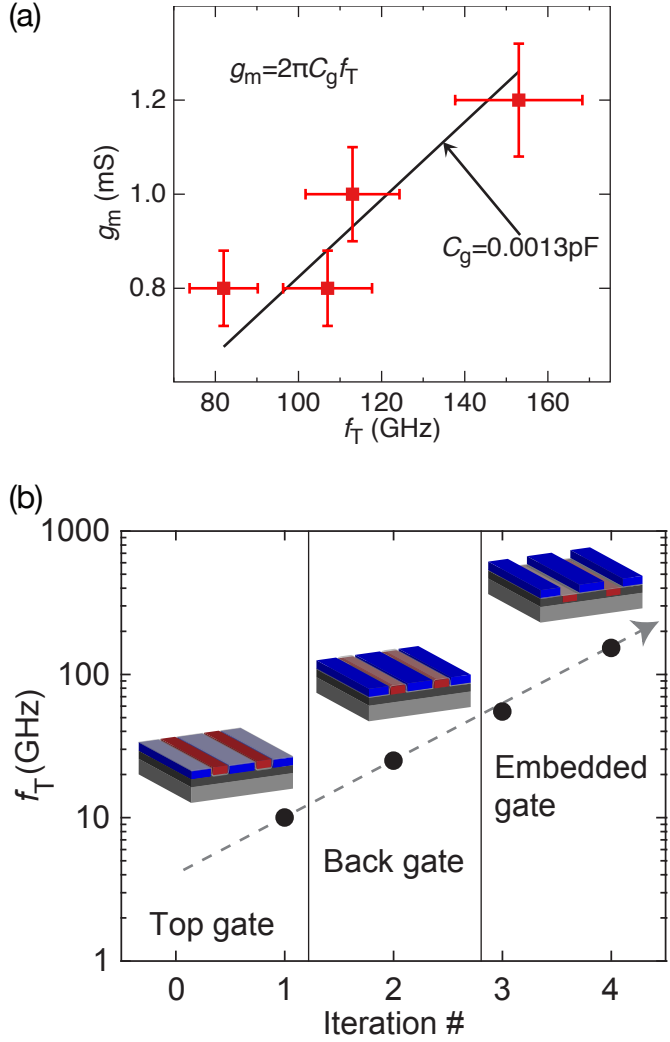

FIG. 4. (a) Plot of the measured transconductance $g_m$ (red squares) as function of the experimental, intrinsic current gain cut-off frequency $f_T$ for four different carbon nanotube array transistors with gate length of $L_g = 100$nm. A linear fit (black line) delivers the experimental value of the gate capacitance $C_g^{exp}$ that is also indicated. (b) Highest intrinsic cut-off frequency $f_T$ achieved (black circles) versus number of experimental iteration. The vertical lines separate the iteration steps that are based on different device (gate) concepts. In the device schematics, the gate electrodes are visualized in red color, source and drain electrodes are shown in blue color.

# Supplementary Material

# High-frequency performance of scaled carbon nanotube array field-effect transistors

Mathias Steiner[1,a], Michael Engel[2,3], Yu-Ming Lin[1], Yanqing Wu[1], Keith Jenkins[1], Damon B. Farmer[1], Jefford J. Humes[4], Nathan L. Yoder[4], Jung-Woo T. Seo[5], Alexander A. Green[5], Mark C. Hersam[5], Ralph Krupke[2,3,6], and Phaedon Avouris[1,b]

[1] *IBM Thomas J. Watson Research Center, Yorktown Heights, New York 10598, USA*

[2] *Institute of Nanotechnology, Karlsruhe Institute of Technology, 76021 Karlsruhe, Germany*

[3] *DFG Center for Functional Nanostructures (CFN), 76028 Karlsruhe, Germany*

[4] *NanoIntegris Inc., Skokie, IL 60077, USA*

[5] *Department of Materials Science and Engineering and Department of Chemistry, Northwestern University, Evanston, IL 60208, USA*

[6] *Institut für Materialwissenschaft, Technische Universität Darmstadt, 64287 Darmstadt, Germany*

---

a) Electronic mail: msteine@us.ibm.com
b) Electronic mail: avouris@us.ibm.com

**Electrical transport measurements**

The DC characterization of the semiconducting carbon nanotube (CNT) array transistors was performed in a probe station (LakeShore FWP6) under $N_2$ flow using a semiconductor parameter analyzer (Agilent B1500A Semiconductor Device Analyzer) and ground-signal-ground coplanar probes (GGB Industries). For high-frequency measurements the probe station was evacuated to a base pressure below $10^{-5}$ mTorr. The high-frequency performance of the transistors was obtained by measuring the scattering (S) parameters using a network analyzer (Agilent E8364C PNA Microwave Network Analyzer). DC voltages at the gate and drain contacts were applied through bias tees using the same semiconductor parameter analyzer.

The intrinsic high-frequency performance was obtained by using a de-embedding procedure based on "open" and "short" reference structures, similar to the method described in references S1 and S2. The "open" reference structure was identical to the active transistor except for the absence of CNTs in the device channel. The short circuit current gain $|h_{21}|$ was determined from the S parameters by using

$$h_{21} = \frac{-2S_{21}}{(1-S_{11})(1+S_{22})+S_{12}S_{21}} \quad . \tag{I}$$

The current gain cut-off frequency $f_T$ is defined as the frequency at which $|h_{21}|$ is equal to 0dB.

The maximum available power gain (MAG) was derived from the S parameters by using

$$MAG = \left|\frac{S_{21}}{S_{12}}\right|\left(K - \sqrt{K^2 - 1}\right) \, . \tag{II}$$

where K is the stability factor expressed by

$$K = \frac{1 + |\Delta|^2 - |S_{11}|^2 - |S_{22}|^2}{2|S_{12}S_{21}|} \quad \text{and} \quad \Delta = S_{11}S_{22} - S_{12}S_{21} \, . \tag{III}$$

The unilateral power gain (U) was determined from the S parameters by using

$$U = \frac{\left|\frac{S_{21}}{S_{12}} - 1\right|^2}{2K\left|\frac{S_{21}}{S_{12}}\right| - 2\,\text{Re}\left(\frac{S_{21}}{S_{12}}\right)} \, . \tag{IV}$$

Both MAG and U were converted into the dB-scale (10 · log |MAG|). The maximum frequency of oscillation $f_{MAX}$ is defined as the frequency at which U (MAG) equals 0dB.